\documentclass[
aps,prl,twocolumn,superscriptaddress,showpacs
 amsmath,amssymb,
]{revtex4-2}

\usepackage{hyperref}
\usepackage{color}
\usepackage{pstool}
\usepackage{units}
\usepackage{graphicx}
\usepackage{amsmath}
\usepackage{bm}

\newcommand{\lambdapump}{\lambda_{\text{pump}}}
\newcommand{\lambdap}{\lambda_{p}}
\newcommand{\lambdaq}{\lambda_{q}}
\newcommand{\lambdar}{\lambda_{r}}
\newcommand{\lambdas}{\lambda_{s}}
\newcommand{\lambdah}{\lambda_{h}}
\newcommand{\omegap}{\omega_{p}}
\newcommand{\omegaq}{\omega_{q}}
\newcommand{\omegar}{\omega_{r}}
\newcommand{\omegas}{\omega_{s}}

\newcommand{\pumpenergy}{\mathcal{E}_{\text{pump}}}
\newcommand{\pcontrolenergy}{\mathcal{E}_{p}}
\newcommand{\qcontrolenergy}{\mathcal{E}_{q}}
\newcommand{\gtwoxc}{g_{XC}^{(2)}}
\newcommand{\gtwoxcbg}{\tilde{g}_{XC}^{(2)}}
\newcommand{\gtwoac}{g_{AC}^{(2)}}
\definecolor{matlabgreen}{rgb}{0,0.5,0}
\EndPreamble

\begin{document}


\title{Toward Deterministic Sources: Photon Generation in a Fiber-Cavity Quantum Memory}

\author{Philip~J. Bustard}
 \email{philip.bustard@nrc-cnrc.gc.ca}
 \homepage{http://quantumtechnology.ca}
\affiliation{National Research Council of Canada, 100 Sussex Drive, Ottawa, Ontario, K1A 0R6, Canada}
\author{Ramy Tannous}
\affiliation{National Research Council of Canada, 100 Sussex Drive, Ottawa, Ontario, K1A 0R6, Canada}
\author{Kent Bonsma-Fisher}
\affiliation{National Research Council of Canada, 100 Sussex Drive, Ottawa, Ontario, K1A 0R6, Canada}
\author{Daniel Poitras}
\affiliation{National Research Council of Canada, 100 Sussex Drive, Ottawa, Ontario, K1A 0R6, Canada}
\author{Cyril Hnatovsky}
\affiliation{National Research Council of Canada, 100 Sussex Drive, Ottawa, Ontario, K1A 0R6, Canada}
\author{Stephen J. Mihailov}
\affiliation{National Research Council of Canada, 100 Sussex Drive, Ottawa, Ontario, K1A 0R6, Canada}
\author{Duncan England}
 \email{duncan.england@nrc-cnrc.gc.ca}
\affiliation{National Research Council of Canada, 100 Sussex Drive, Ottawa, Ontario, K1A 0R6, Canada}
\author{Benjamin~J. Sussman}
\affiliation{National Research Council of Canada, 100 Sussex Drive, Ottawa, Ontario, K1A 0R6, Canada}
\affiliation{Department of Physics, University of Ottawa, Ottawa, Ontario, K1N 6N5, Canada}
\date{\today}





\begin{abstract}
We demonstrate generation of photons within a fiber cavity quantum memory, followed by later on-demand readout. Signal photons are generated by spontaneous four-wave mixing in a fiber cavity comprised of a birefringent fiber with dichroic reflective end facets. Detection of the partner herald photon indicates the creation of the stored signal photon. After a delay, the signal photon is switched out of resonance with the fiber cavity by intracavity frequency translation using Bragg scattering four-wave mixing, driven by ancillary control pulses. We measure sub-Poissonian statistics in the output signal mode, with $\gtwoac=0.54(1)$ in the first readout bin and a readout frequency translation efficiency of $\approx80\%$. The 1/e memory lifetime is $\approx67$ cavity cycles, or \unit[1.68]{$\mu$s}. In an alternate fiber cavity, we show a strategy for noise reduction and measure $\gtwoac=0.068(10)$ after one cavity cycle.
\end{abstract}

\maketitle


Photons are a key resource in many quantum technology designs~\cite{nphoton.3.687}, including quantum computing~\cite{Nature.409.46,RevModPhys.79.135}, quantum sensing~\cite{Degen2017}, and quantum communication~\cite{RevModPhys.74.145}. These technologies provide much of the impetus for current quantum photonics research, where exquisite control of photon properties is the goal. Among the various spatial, spectral, polarization, and temporal attributes to be controlled, the temporal interval in which a photon exists is often of critical importance. When multiple photons are required to interact in an interferometric network, for example, they must arrive with precise timing control~\cite{Science.339.794,Science.339.798,ZhongScience2020}. Parametric photon pair sources were used for these seminal demonstrations in photonic quantum processing. The probabilistic operation of photon pair sources is one of the factors limiting efforts toward scalable photonic quantum computing~\cite{Nature.409.46,RevModPhys.79.135}. Methods to produce photons deterministically, or `on demand', have therefore been the source of much research in recent years. Single emitter sources, such as quantum dots~\cite{Loredo2016}, offer one approach; another is multiplexing, where independent probabilistic sources are combined with event-driven switching to create a quasi-deterministic source~\cite{Migdall2002,Pittman2002,MeyerScott2020}.

A variety of degrees of freedom have been used to demonstrate improved success probability from a photon pair source, including multiplexing of distinct spatial modes~\cite{Ma2011,Collins2013,Xiong2013,Meany2014,FrancisJones2016,Kiyohara2016,Mendoza2016}, frequency modes~\cite{Puigibert2017,Joshi2018,Hiemstra2020}, and temporal modes~\cite{Kaneda2015,Xiong2016,Hoggarth2017,Kaneda2017,Kaneda2019,Yoshikawa2013,Bouillard2019}. In each case, photon loss counteracts the benefits gained from a switch operating with independent sources. Effective solutions should ideally combine efficient, low-loss, low noise switching with realistic resource scaling requirements. Temporal multiplexing is attractive because fewer resources are required, when compared to alternatives. The idea is to combine a storage cavity~\cite{Pittman2002}, or quantum memory~\cite{PhysRevLett.110.133601}, with an optical switch. Memories suited to this strategy will have a high acceptance bandwidth relative to their lifetime (a high ``time-bandwidth product'') so that many generation attempts can be completed in close succession before storage and subsequent read out~\cite{PhysRevLett.110.133601}. With careful design and optimized components, promising results have been attained in free space cavities, including temporal multiplexing to prepare single-photon~\cite{Kaneda2015,Kaneda2019}, multi-photon~\cite{Kaneda2017,Engelkemeier2021}, and entangled~\cite{MeyerScott2022}, states. Opto-electronic switching is attractive because of high efficiency and low noise. Nonetheless, loss due to diffraction and surface reflections serves to limit the overall efficiency of free space cavities, both intracavity and when transferring a pulse into the cavity.

Optical fiber technologies offer complementary benefits to their free-space counterparts. Single-mode fibers offer flexible routing, low-loss transmission, and tight optical confinement. Fibers are often used as convenient, low-loss, fixed-time-delay ``buffers'' to compensate for electronic latency in multiplexing experiments~\cite{Xiong2016,FrancisJones2016,Kaneda2017,Kaneda2019,Lee2019,MeyerScott2022}. For a variable-delay quantum memory based on a fiber-cavity design~\cite{Leung2006}, fiber integrated switches might attain high efficiencies, but their insertion losses~\cite{Margulis2011} severely limit the cavity lifetime. Recently, we demonstrated that fiber-cavity storage with intracavity frequency translation (FC SWIFT) allows high efficiency memory operation at the quantum level~\cite{Bustard2022}. We later explored a more narrowband version of FC SWIFT in the telecoms C-band, using fiber Bragg gratings as reflective elements, instead of dielectric coatings~\cite{BonsmaFisher2023}. Here, we demonstrate the heralded generation of broadband signal photons within a fiber cavity, where they are temporarily stored before later on-demand release. This approach eliminates the routing losses associated with sources external to a quantum memory and uses nonlinear optical switching to avoid the insertion losses of an electro-optic switch. We note that the generation of a signal quantum within a memory is recognized as a major benefit of the DLCZ scheme~\cite{Duan2001}, typically used in atomic vapours; this scheme has recently been used in multiplexing experiments with MHz~\cite{Heller2020,Dideriksen2021} and GHz~\cite{Zhang2022} bandwidth photons. Our alternative memory-integrated source offers a promising route to deterministic generation of THz-bandwidth single-photon and multi-photon states.

\begin{figure*}
\begin{center}
\includegraphics[width=2\columnwidth]{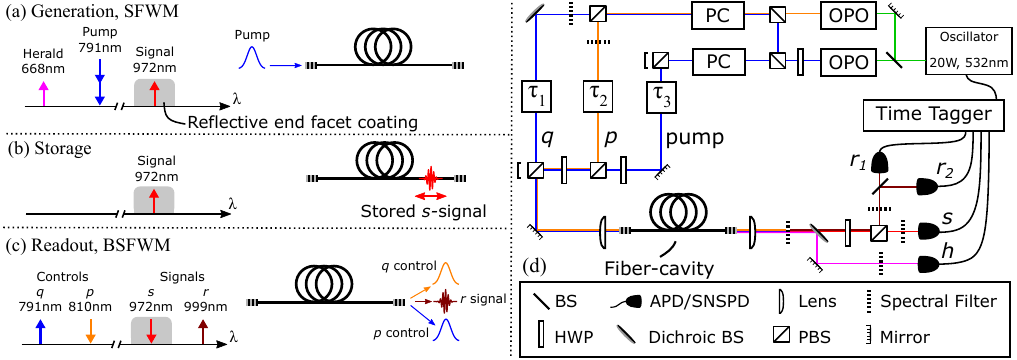}
\end{center}
\caption{(a) Generation step: a bright pump pulse generates a signal and herald pair by SFWM, with the signal trapped by a dichroic reflective end facet coating. (b) Storage step: the signal photon is trapped in the cavity. (c) Readout step: bright control pulses translate the signal $s\rightarrow r$ by BSFWM. (d) Experiment layout. BS: beam splitter; PBS: polarizing beam splitter; HWP: half-wave plate.\label{fig:experiment}}
\end{figure*}

Our protocol is shown in Fig.~\ref{fig:experiment}. 
The birefringent fiber is a linear cavity, with end facets each coated with a dielectric stack to have high reflectivity at $\lambdas$ and low reflectivity at other wavelengths. In the \textsl{generation} step, a pump pulse propagating on the fiber slow axis generates signal and herald photon pairs on the fast axis by spontaneous four-wave mixing~\cite{OptExpress.17.23589}. The herald photon with wavelength $\lambdah$ exits the fiber cavity and is detected, indicating that a signal photon has been created. The signal photon, wavelength $\lambdas$, is reflected by the dielectric stack and stored within the cavity. After a delay, two intense control pulses, wavelengths $\lambdap,\lambdaq$ are focused into the fiber cavity, timed to overlap with the trapped signal photon. In this \textsl{read} step, they induce frequency translation of the signal to $\omegar=\omegas-(\omegaq-\omegap)$, where $\omega_i=2\pi c/\lambda_i$, by Bragg scattering four-wave mixing (BSFWM). The end facet coating has a high transmission at $\omegar$ so that the signal exits the fiber cavity and is available for use. 

Our integrated source approach eliminates the losses associated with mapping photons into the cavity and raises the prospect of low loss temporal multiplexing. In single-mode fiber, photons are generated with a spatial mode that is well-suited to interface with other fiber optic elements. The use of polarization-maintaining (PM) birefringent fiber ensures a fixed polarization relationship between the signal and controls across all storage times~\cite{BonsmaFisher2023}. Birefringent fiber also enables phase matching of SFWM~\cite{OptExpress.17.23589} and BSFWM~\cite{OptExpress.26.17145} across a broad spectral range. Since the transmission spectra of optical coatings can be designed, this protocol may thus be operated at a wide variety of wavelengths. 

The sources for the experiment are two synchronously pumped optical parametric oscillators (OPOs), each delivering $\sim\unit[3]{W}$ of tunable near-infrared pulses of duration $\approx$\unit[13]{ps} at \unit[80.1]{MHz} repetition rate. Control and pump pulses are picked from the OPO outputs using Pockels cells, allowing adjustment of the relative delay between the pump and control pulses. 

We fabricated a cavity from a polarization-maintaining (PM) fiber (HB800) of length $L\approx\unit[2468]{mm}$, such that the cavity frequency $\nu_{\text{c}}\approx\unit[40.05]{MHz}$. Dichroic dielectric coatings were deposited on the end facets of the fiber. The coatings were designed to obtain high reflectivity at $\sim\unit[971]{nm}$, with low reflectivity at other specific wavelengths (see Fig.~\ref{fig:experiment}). The fiber is placed under tension, allowing adjustment of the fiber cavity length relative to the primary laser cavity length. Photon pairs of a herald at $\lambdah=\unit[667.6]{nm}$ and a signal at $\lambdas=\unit[971.5]{nm}$ are generated on the fiber fast axis by SFWM, driven by the pump on the slow axis at $\lambdapump=\unit[791.4]{nm}$. Herald photons exit the fiber where their detection by an avalanche photodetector (APD) verifies the creation of a signal photon trapped in the cavity. After a delay, control pulses at $\lambdaq=\unit[791.4]{nm}$ and $\lambdap=\unit[809.7]{nm}$ are synchronously injected to the fiber on the fast and slow axes, respectively. Timed to overlap with the stored signal pulse, they drive BSFWM with the signal, frequency translating it to $\lambdar=\unit[999.2]{nm}$ on the slow axis. Photons exiting the fiber cavity in the herald, trapped signal, and output signal modes are separated by polarization and spectral filtering before collection in single mode fibers. Photons in the output signal mode are partitioned at a fiber beam splitter and detected with two superconducting nanowire single photon detectors (SNSPDs). We measure and record detection events in coincidence with the Pockels cell trigger clock. The clock rate for the experiment was \unit[76.8]{kHz}.

\begin{figure}
\begin{center}
\includegraphics[trim=115 120 158 400,clip,width=\columnwidth]{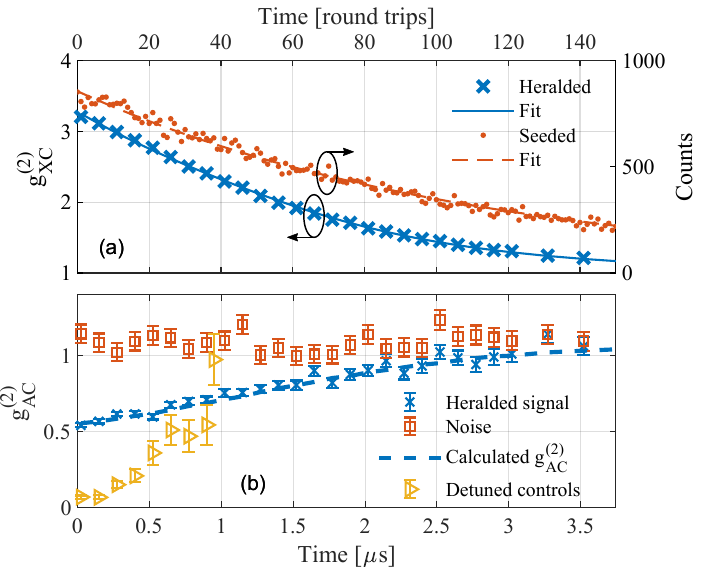} 
\end{center}
\caption{(a, right axis) Seeded FWM ring-down of \unit[971]{nm} signal stored in the cavity (dots, red) with exponential fit (dashed line, red) of $1/e$ lifetime \unit[111(1)]{cycles}. (a, left axis) Two-mode cross correlation between the herald and output signal modes as a function of read out delay (blue crosses) and a model fit (solid blue line). (b) Auto-correlation $g^{(2)}_{AC}$ of the heralded output signal (blue crosses), noise (red squares), and expected auto-correlation (blue dashed line) as a function of read out delay time. The heralded auto-correlation for an alternate cavity with independent pump-control tuning (yellow triangles) shows improved nonclassical statistics. \label{fig:readoutScan}}
\end{figure}

For convenience, the experiment can be run in ``bright pulse mode": we use a continuous wave external cavity diode laser (ECDL) to seed the herald mode, generating bright signal pulses by four-wave mixing with the pump pulse~\cite{GarayPalmett2023}. This allows measurements on the stored signal light. Stored signal pulses gradually leak from the cavity due to the imperfect coating reflectivity. Recording detection events in the stored signal mode relative to the experiment clock, we measure a cavity ring-down with a $1/e$ lifetime of $111(1)\tau_c$, for a cavity cycle time of $\tau_c=\unit[24.96]{ns}$ (see Fig.~\ref{fig:readoutScan}(a)).

We first evaluate the properties of the SFWM photon generation without applying read-out control pulses. All results quoted are based on raw count rates, except where background subtraction is stated, or indicated by a tilde. The seeded FWM signal spectrum from the fiber cavity has a bandwidth of \unit[2.2]{nm}. The high reflectivity coating results in a low count rate for the stored signal, making it difficult to measure the heralded signal auto-correlation. Instead, we measure the two-mode cross-correlation between the herald and $s$-signal photons that immediately exit the fiber cavity. For $\pumpenergy=\unit[0.7]{nJ}$, we measure a raw value of $\gtwoxc(h,s)=16.0(1.6)$ and background-subtracted value of $\gtwoxcbg(h,s)=26(3)$, after correction for accidental dark counts. These values are each far in excess of the classical bound of $\gtwoxc=2$, indicating that we have created a nonclassical state within the fiber cavity. A Klyshko measurement~\cite{Klyshko1980} of the herald collection efficiency as detected yields $\eta_h=0.06(1)$. Using ECDL light exiting the fiber cavity, we measure a collection efficiency of 0.51(2), at the herald wavelength, excluding detection. The discrepancy arises because of the APD efficiency ($\sim0.7$), and because the non-zero reflectivity of the coating at the herald wavelength ($\mathcal{R}_h=0.66$) temporarily traps some of the herald photons in the fiber. In what follows, we only consider coincidence measurements where the herald photon immediately exits the cavity after generation. We have measured coating reflectivities $\mathcal{R}\lesssim0.05$, at wavelengths close to $\lambda_h$, suggesting that a $\times2.8$ increase in herald collection efficiency is possible with an alternative coating.

We measure the retrieval of signal photons from the cavity by applying control pulses, timed to overlap with the stored signal pulse. The $p$-and $q$-control pulse energies are $\pcontrolenergy=\unit[6.9]{nJ}$ and $\qcontrolenergy=\unit[8.6
]{nJ}$, respectively. Figure~\ref{fig:readoutScan}(a) shows the two-mode herald-signal cross-correlation $\gtwoxc(h,r)$ and Fig.~\ref{fig:readoutScan}(b) shows the heralded output signal auto-correlation $\gtwoac(r_1,r_2|h)$ as a function of readout delay, measured at intervals of \unit[5]{cycles}. Data was acquired over a 17-hour period, during which time the passive cavity stability was sufficient to maintain readout synchronization over $\approx100$\,cycles. The cross-correlation obtains a maximum value of $\gtwoxc(h,r;T=1)=3.21(1)$, where $T$ is the generation-readout delay in cavity cycles. This non-classical signal-herald correlation has been maintained after frequency translation; however, the addition of noise photons in the readout interaction has reduced the degree of correlation. The $1/e$ memory lifetime is $\approx67$ cycles, reduced from the cavity ring-down measurement. The decreasing signal is no longer a simple exponential. We attribute this to the imperfect matching of the fiber cavity cycle time, compared to the OPO cycle time, and the effect of dispersion chirping the signal as it propagates. A simple model allowing for the cavity loss, the estimated second order dispersion ($\psi_2=$\unit[0.05]{ps$^2$/cycle}), and the reduced conversion efficiency from signal-control walk-off, gives good agreement with the data, with $R^2=0.9992$, for a cavity-mismatch walk-off of \unit[0.09]{ps/cycle} and an inferred cavity lifetime of \unit[78]{cycles}. We attribute the discrepancy between the inferred lifetime and the ring down measurement to differences in the seeded and spontaneous signal spectra. 

The heralded output signal auto-correlation obtains a minimum value of $\gtwoac(r_1,r_2|h;T=1)=0.54(1)$, indicating that the retrieved signal pulse exhibits sub-Poissonian statistics. The unheralded noise has an average value of $g^{(2)}_{\text{AC,noise}}(r_1,r_2)=1.09(1)$, consistent with a multi-mode sum of thermal contributions. Over the succeeding \unit[80]{cycles}, or \unit[3]{$\mu$s} of delay, the heralded auto-correlation gradually tends towards equality with the noise measurement for the controls only such that $\gtwoac(r_1,r_2|h;\,T>80)\approx g^{(2)}_{\text{AC,noise}}(r_1,r_2)$. A calculation of the expected output signal auto-correlation, modelled as an incoherent mixture of the stored signal and the noise, each with known $\gtwoac$, shows good agreement with the measured values (Fig.~\ref{fig:readoutScan}(b), blue dashed curve)~\cite{Goldschmidt2013,NJP.17.043006}.

\begin{figure}
\begin{center}
\includegraphics[trim=115 113 158 458,clip,width=\columnwidth]{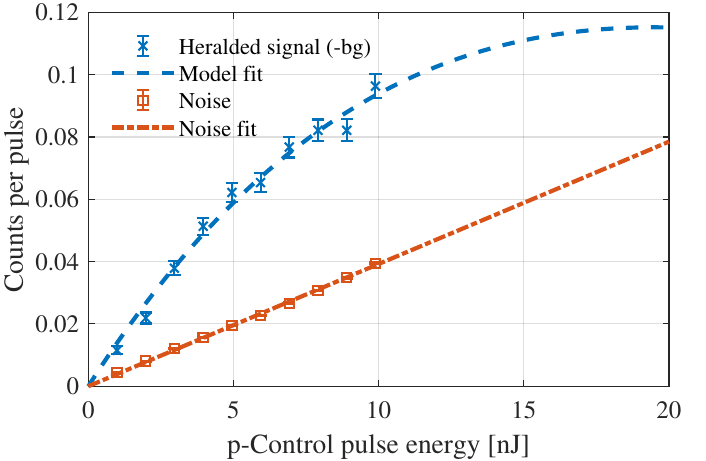}
\end{center}
\caption{Background-subtracted heralded output signal count rate (blue crosses) and unheralded noise count rate (red squares) as a function of p-control pulse energy for fixed q-control pulse energy. Analytic model fit to signal (dashed blue line) and linear fit to noise (dash-dot red line). See text for details. \label{fig:pPowerScan}}
\end{figure}

In the $T=1$ bin, we measured maximum raw count rates of $R_h=\unit[474(15)]{cps}$, $R_r=\unit[3405(13)]{cps}$, $R_{h,r}=\unit[63(2)]{cps}$, and $R_{h,r1,r2}\approx\unit[0.8]{cps}$ for the herald, $r$-signal, $(h,r)$ two-fold coincidences, and $(h,r_1,r_2)$ three-fold coincidences when operating the experiment at \unit[76.8]{kHz}. Figure~\ref{fig:pPowerScan} shows a measurement of the background-subtracted heralded signal rate per pulse, or heralding efficiency (blue crosses), compared to the unconditional noise rate (red squares), as a function of p-control pulse energy. 
The maximum heralding efficiency is $\tilde{p}(r|h)=0.096(4)$, corresponding to a signal generation rate of $\tilde{R}_{r|h}=46(3)$\,s$^{-1}$. We measure a coating reflectivity of $\mathcal{R}_r=0.45$; improved coating design so that $\mathcal{R}_r\lesssim0.05$ would improve the heralding efficiency by $\times1.7$. Noise photons in the output bin are mostly generated by the $p$-control pulse, polarized on the slow axis with the output signal photons. Figure~\ref{fig:pPowerScan} shows that the noise increases linearly as a function of p-control pulse energy for fixed q-control pulse energy, consistent with spontaneous Raman scattering as the noise source. 

The heralded output signal rate varies approximately linearly at low p-control energies (Fig.~\ref{fig:pPowerScan}) before starting to turn over due to stored signal depletion. For optimal BSFWM phase matching, using Gaussian control pulses, the output field $a_r(t)$ is given by~\cite{OptExpress.26.17145} ${a_r(t)=ie^{4i\xi(t)}\sin\left[\xi(t)\right]a_s(t)}$, and,
\begin{equation}
\xi(t)=\frac{\gamma\sqrt{\pcontrolenergy\qcontrolenergy}}{3\beta}\left[\text{erf}\left(\frac{t}{\tau}+\frac{\zeta}{2}\right)-\text{erf}\left(\frac{t}{\tau}-\frac{\zeta}{2}\right)\right],
\end{equation}
where $t$ is time in the signal frame, $\gamma$ is the nonlinear coefficient, and $\tau=\tau_{\text{FWHM}}\sqrt{4\ln{2}}$ where $\tau_{\text{FWHM}}=\unit[13.5]{ps}$ is the full-width at half-maximum duration of the control pulses. For the estimated group velocity walk-off of $\beta=\unit[13.5]{ps/m}$, we have $\zeta=\beta L/\tau\approx4.1$. Allowing a free scaling factor for the imperfect coating transmission, collection efficiency, and detector efficiency, we fit for the unknown nonlinear coefficient $\gamma$ and find good agreement with the background-subtracted heralded signal photon rate (Fig.~\ref{fig:pPowerScan} dashed blue curve). The internal conversion efficiency $\eta_{\text{BSFWM}}=0.8$ at maximum power agrees well with the $\approx80\%$ stored signal depletion when operating in bright pulse mode. Based on this fit, an increase of approximately 40\% in both control pulse energies would be required to attain the maximum read out efficiency of $\approx98\%$ in the current configuration. This would increase the heralding efficiency by $\times1.22$.


Given the deleterious impact of the control-induced noise on the output signal photon statistics, we investigated a potential mitigation strategy. We locked a Ti:sapphire oscillator to the OPO system and used the oscillator to pump the SFWM generation step. We operated the memory in an alternate fiber cavity sample with a coating that permitted the controls to be blue-detuned from the pump by \unit[16.7]{THz}. Figure~\ref{fig:readoutScan}(b) shows the heralded output signal auto-correlation as a function of pump-read pulse delay. The lifetime of $\approx$\unit[12]{cycles} is reduced, compared to the primary sample; however, the photon statistics are significantly improved due to the reduced number of noise photons in the signal mode. We measured ${\gtwoac(r_1,r_2|h;T=1)=0.068(10)}$ after one cavity cycle, demonstrating the potential for improved nonclassical statistics in future demonstrations of this scheme.

The present implementation is limited by several factors. Firstly, imperfect coating transmission at both the herald wavelength and output signal wavelength limit the herald collection efficiency and the heralding efficiency, respectively. Design improvements are likely to deliver better herald and output signal transmission coefficients while retaining a relatively long cavity lifetime, so that temporal multiplexing is a realistic goal. 
Secondly, imperfect matching of the OPO and fiber-cavity frequencies reduced the memory lifetime to 67\,cycles.
The fiber tension required to match the cavity lifetime also affected the phase matching for SFWM and BSFWM. Our results represent a balance of matching these competing factors against the fixed transmission spectrum of the coating. Closer matching of the untensioned fiber cavity frequency to the laser repetition rate, or adjustable rate lasers, would mitigate this problem. Finally, the noise photons introduced by the bright control pulses can be reduced. As we demonstrated in an alternate fiber cavity, independent tuning of the control wavelengths relative to the pump wavelength can increase the Stokes shift between the control and signal wavelengths, and thus reduce the Raman scattering noise. Alternatively, since polarization maintaining fibers permit broadband phase matching~\cite{OptExpress.17.23589,OptExpress.26.17145}, setting the signal wavelength on the anti-Stokes side of the controls would further reduce the noise.

We created signal photons with bandwidth $\delta\omegas=\unit[0.6]{THz}$ by SFWM in an end-facet-coated fiber cavity, subsequently using BSFWM to frequency translate the signal photons out of resonance with the cavity. We demonstrated sub-Poissonian statistics in 80 independent time bins, raising the possibility of temporal multiplexing across many independent sources~\cite{MeyerScott2020}. Temporal multiplexing of up to $K=40$ independent sources has been demonstrated to offer $\times9.7(5)$ improvement in photon generation efficiency, using a free-space cavity with a 1/e lifetime of 83 cycles~\cite{Kaneda2019}. In this demonstration, we achieved a 1/e memory lifetime of 67\,cycles and a ring-down cavity lifetime of 111\,cycles. Given the additional benefits from the integrated SFWM source and fiber mode compatibility, we expect our scheme to offer comparable efficiency benefits for multiplexed generation of single- and multi-photon Fock states.

 \begin{acknowledgements}
 We are grateful for discussions with Neil Graddage, Khabat Heshami, Aaron Goldberg, Fr\'ed\'eric Bouchard, Kate Fenwick, Guillaume Thekkadath, and Yingwen Zhang. We also thank Huimin Ding, Denis Guay, Rune Lausten, and Doug Moffatt for technical assistance.
 \end{acknowledgements}






%

\end{document}